\documentclass[a4paper]{jpconf}
\usepackage{graphicx}
\usepackage{pifont}
\usepackage[numbers]{natbib}
\usepackage{hyperref}
\usepackage{verbatim}
\usepackage{amsmath}
\hypersetup{
colorlinks = true,
urlcolor   = blue,
linkcolor  = blue,
citecolor  = blue
}
\usepackage{comment}
\begin{document}
\title{TRIM Simulations Tool for $\mu^+$ Stopping Fraction in Hydrostatic Pressure Cells}

\author{Frank Elson$^{1,*}$, Debarchan Das$^2$, Gediminas Simutis$^{3,4}$, Ola Kenji Forslund$^4$, Ugne Miniotaite$^1$, Rasmus Palm$^1$, Yasmine Sassa$^4$, Jonas Weissenrieder$^1$, and Martin M{\aa}nsson$^{1,+}$}
\address{$^1$Department of Applied Physics, KTH Royal Institute of Technology, SE-106 91 Stockholm, Sweden}
\address{$^2$Laboratory for Muon Spin Spectroscopy, Paul Scherrer Institut, CH-5232 Villigen PSI, Switzerland}
\address{$^3$Laboratory for Neutron and Muon Instrumentation, Paul Scherrer Institut, CH-5232 Villigen PSI, Switzerland}
\address{$^4$Department of Physics, Chalmers University of Technology, Göteborg, SE-412 96, Sweden}

\ead{$^{*}$elson@kth.se, $^{+}$condmat@kth.se}

\begin{abstract}
For quantum systems or materials, a common procedure for probing their behaviour is to tune electronic/magnetic properties using external parameters, e.g. temperature, magnetic field or pressure. Pressure application as an external stimuli is a widely used tool, where the sample in question is inserted into a pressure cell providing a hydrostatic pressure condition. Such device causes some practical problems when using in Muon Spin Rotation/Relaxation ($\mu^+$SR) experiments as a large proportion of the muons will be implanted in the pressure cell rather than in the sample, resulting in a higher background signal. This issue gets further amplified when the temperature dependent response from the sample is much smaller than that of the pressure cell,which may cause the sample response to be lost in the background and cause difficulties in aligning the sample within the beam.  To tackle this issue, we have used pySRIM \cite{pysrim} to construct a practical and helpful simulation tool for calculating muon stopping fractions, specifically for the pressure cell setup at the $\mu$E1 beamline using the GPD spectrometer at the Paul Scherrer Institute, with the use of TRIM simulations. The program is used to estimate the number of muon stopping in both the sample and the pressure cell at a given momentum. The simulation tool is programmed into a GUI, making it accessible to user to approximate prior to their experiments at GPD what fractions will belong to the sample and the pressure cell in their fitting procedure.
\end{abstract}

\section{Introduction}
High pressure studies using muon spin resonance ($\mu^+$SR) has become very popular as a 'clean' approach for probing and tuning material properties, and especially within the field of magnetic materials and quantum criticality \cite{Forslund_2019,Sugiyama_2020,Sugiyama_2021}. The purpose of this is to investigate the phase diagram of a given material as a function of an external perturbation (i.e. pressure but also e.g. temperature and magnetic field). Through such an approach, it is possible to shed light onto a plethora of complex phenomena, such as unconventional superconductivity \cite{SuperCon}, quantum spin liquids \cite{Spin_Liquid} or frustrated magnetism \cite{Frustrated}. The combination of pressure and the $\mu^+$SR technique, in which a microscopic understanding of the ground state of these quantum materials can be obtained, provides a uniquely powerful tool to get insight into the intrinsic properties of the material. 
However, to apply hydrostatic pressure (e.g. at GPD, which will be the spectrometer used as the example throughout this paper), the samle needs to be compressed into a pellet and inserted into the pressure cell made of either MP45N or Cu-Be. Before starting any measurements, the muon beam and its momentum need to be adjusted to guarantee the maximal signal from the sample. With surface muon beamlines (e.g. GPS at PSI), the energy of the muons is usually 5 - 40 MeV/c. This is enough to penetrate a thin layer of mylar foil or capton tape which is used to mount the sample to the sample holder. However, this energy is not enough to penetrate the thick layers of metal that make up a pressure cell. For this, decay muon beamlines are needed, which can provide muons in the energy range of 40 - 125 MeV/c. For the surface muon case, alignment is simple as there is essentially only the sample signal to detect. When the pressure cell is introduced, it gets slightly more difficult as there is now another material in very close proximity to the sample that will give a significant signal. This is not an issue when the sample gives a very strong response, as the sample signal will be much greater than the pressure cell signal. Consequently, determining the optimal muon momentum for alignment is not very difficult. However, if the sample response is weak, then it may be difficult to distinguish the sample contributions from the pressure cell (as mentioned in Ref~\cite{GPD_Details}). Additionally, there are different types of pressure cells (MP35N, CuBe and a combination of the two) with different signals. Further, each individual pressure cell made from same material are in fact also slightly different. As a result, there are several cases where it is incredibly difficult to extract the sample response from the total signal.

One way around this is to first employ an indirect alignment using another sample with a strong response (but same pressure cell), and then using this alignment result also for the sample of interest (with weak signal). However, this can be a problem as slight differences in the sample densities can drastically change the stopping fraction of muons in the sample (a higher density means less muons pass through the sample). This then translates into issues in the fitting process as you need to have a correct estimate for the signal fraction from the sample. Consequently, if your assumed stopping fraction is incorrect, the obtain fitting results do not completely reflect the intrinsic physical properties and behaviour of the sample.

In this paper, we present an efficient and user friendly simulation method based using TRIM \cite{TRIM} combined with the pySRIM python module \cite{pysrim}. In such approach, it is possible to simulate, with higher level of accuracy, the number of muons that will be stopping in the pressure cell and the sample, respectively, for any given muon momentum input. This is useful as a tool for user to employ both before and after experiments. This is because it will both give the user an idea on feasibility of their proposed experiment (in the respect of how easy it will be to see their sample signal) and also for supporting the data analysis when fitting the percentages of the contributions from each fit function (pressure cell and sample).

\section{Basis of Simulation}
\subsection{Beam setup}
The GUI software is a python package that determines the stopping fractions in a beamline setup. The software runs TRIM calculations in the background and presents the results as early interpretable figures and stopping percentages.

\begin{figure}[ht]
  \begin{center}
    \includegraphics[scale=0.55]{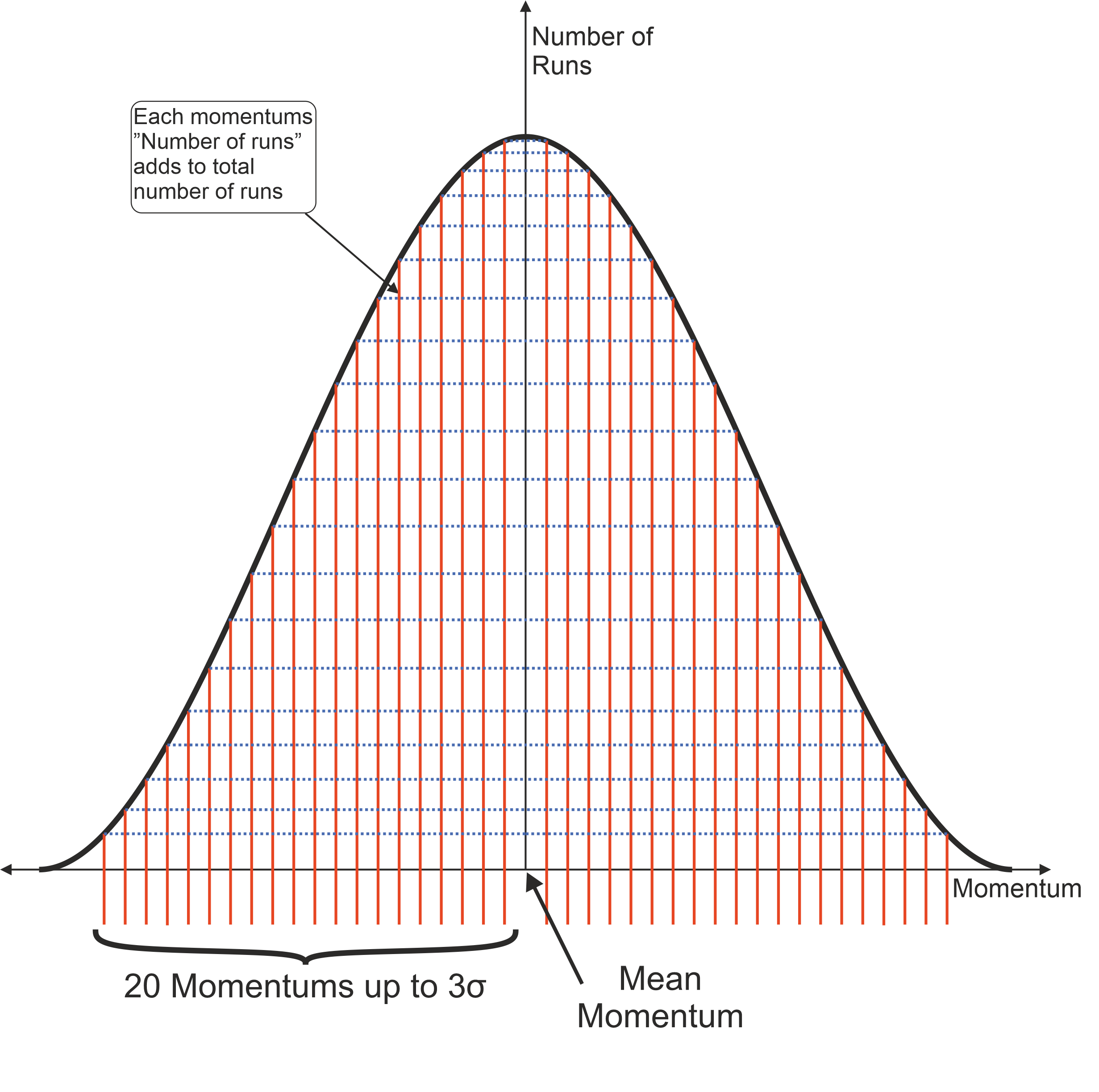}
  \end{center}
  \caption{A graph to show how the number of runs at each energy is decided. The height at each momentum is weighted such that it fits the Gaussian shape, and when summed with the other heights it equals a value very close to the number of simulations input into the software.}
  \label{Num_Sims}
\end{figure}

There are three key aspects that make this software more accurate for modelling stopping fractions compared to running the standard TRIM simulations:
\begin{enumerate}
    \item The software utilises a Gaussian distribution of momentums/energies.
    \item The input beam can be collimated to any area. 
    \item There is a random small angular divergence on each muon as each muon will not be completely parallel to the beam.
\end{enumerate}
Setting up the energy distribution in the beam first requires an input for the percentage momentum acceptance. This represents the full width half maximum of the momentum distribution. At the $\mu$E1 beam line, the momentum acceptance is 3$\%$ \cite{GPD_Details}, meaning a mean momentum of 95~MeV/c will give a full width half maximum of 2.85~MeV/c. Thereafter a momentum range of 3$\sigma$ is selected on both sides of the mean momentum. This is the initial input to the software that then samples 20 discrete momenta within the $\pm$3$\sigma$ range. The number of runs at each of these momenta is then weighted so that the sum of these values gives a result as close to input number of runs, whilst also following a Gaussian distribution, as seen in Fig.~\Ref{Num_Sims}. The scaling value is equal to the input number of runs divided by the sum of the selected $y$-values on the Gaussian. This means that the total number of simulations will either be less or more than the given total number of simulations. However, it is never by such a large value that it becomes an issue. For 1000 runs (where one run is one muon simulated) at a momentum of 95~MeV and 3$\%$ momentum acceptance, 998 runs were completed, and for 10000 runs at the same energy 1006 runs were completed. The momentum values then need to be converted into their energy value, as TRIM only takes energy values as the input. We then create a list of corresponding energy values using the following equation:

\begin{equation}
    E = \sqrt{\left(pc\right)^2 + \left(m_0c^2\right)^2} - m_0c^2
\end{equation}
where $p$ is the momentum, $c$ is the speed of light, and $m_0$ is the rest mass of the muons $\left( m_0 = 105.658~MeV/c^2 \right)$.

Each muons initial conditions (energy, position and angular divergence) is written into the {\tt trim.dat} file, which is provided when the SRIM software is installed. The initial positions are decided according to the collimation at GPD, 4x10mm$^2$ \cite{GPD_Details}. Therefore the $y$ and $z$ positions are randomly sampled according to a Gaussian distribution with full width half maximums of 2mm and 5mm respectively, to match beam size measurements taken at GPD. A small angular divergence is added in also (-5$^\circ$ to 5$^\circ$ along all three axes) using {\tt numpy.random.uniform} in python. This in total created more accurate representation of the beam setup at GPD.

\subsection{Modelling the GPD setup}
TRIM only allows to form layers of materials, and there is no possibility to introduce either overlapping layers, or materials that are completely surrounded by other materials. In other words, we had to model the GPD setup as accurately as possible using only stacked individual layers of different thickness and density. In the current setup for the simulations we primarily target only the Variox-Heliox cryostat used at GPD (but different setups are planned to be implemented later). The Variox is the first layer the muons see, consisting off 0.1~mm Mylar, 0.2~mm Aluminium Nitrogen shield and 0.8~mm Copper. Further, the Heliox consists of 2~mm thick Brass.

After penetrating the cryostat layers, the muon will reach the pressure cell and finally the sample. At GPD, 3 types of pressure cells are used, \textcircled{\small{1}} made entirely of MP35N alloy (Nickel - 35 , Cobalt - 35, Cromium - 20 , Molybdenum - 10, with a density of 8.43 g/cm$^3$), \textcircled{\small{2}} made entirely of CuBe alloy (Berylium - 3, Copper - 97, with a density of 8.25 g/cm$^3$) and finally \textcircled{\small{3}} a double-wall cell consisting of an inner chamber of CuBe alloy and the outer part of the pressure cell made of MP35N alloy. All three cells have an outer diameter 24~mm along with a sample space with diameter 6~mm and a height 12~mm. For the double-wall cell \textcircled{\small{3}}, the inner CuBe chamber has an outer diameter 12~mm \cite{PC_Details}. All cells are in the simulation represented as slabs of: 'pressure cell - sample - pressure cell'.

\subsection{Obtained Results}
From the simulations, the $(x,y,z)$ position of every muon is collected. The only exception to this is when the energy of the muon is so high that it passes straight through the entire setup (all layers) and exits on the other side. In this case, the muons stopping position is not saved (as it is out of the range of calculation). 
%However, this will be easily seen on the plots (as very few points plotted at that energy) and also this energy scale would be useless to the goal of the simulations, and so this is not an issue. 
By collecting the coordinates of each muon, the stopping fraction for the number of muons inside the sample can be calculated. The $(x,z)$ plane cuts across the circular area of the sample/pressure cell, and the $y$ direction describes the height of the sample/pressure cell. Choosing the origin to be the sample center, the absolute value of the $x$ and $z$ coordinates of the moun is extracted, and evaluate if the square root of the sum of the square of such coordinate ($\sqrt{x^2 + z^2}$) is less than the radius of the sample area (3~mm). If this condition is met and the $y$ coordinate is within -6 and 6 (for a 12~mm height with $z=0$ as the middle oft the sample) then the muon is counted as being inside the sample. This number divided by the total number of simulations will finally give the stopping fraction.

There are a few assumptions made for this simulation. As mentioned earlier, we can only simulate the GPD setup as "slabs" of material. Therefore, the sample height is not introduced in the simulations, and only comes in when working out the sample stopping fraction. This means we have taken small cylindrical region of the sample slab as the "sample region", and all around this region as counting as the sample. This is a small issue due to most samples being put into the beam having a lower density than the pressure cell, meaning that the muons can move more freely in $zy$ plane in the sample slab than they would in the real setup at GPD. However, it is estimated that the extra distance travelled by the muons in this region will only have minuscule effect on the accuracy of the simulations. 

To 'field test' the software We have also compared the simulations to the alignment results on our previous GPD measurement of the CrCl$_3$ material, which used the same setup as the simulations. The stopping fractions were calculated from the asymmetry of the sample in the fit, by taking the total possible asymmetry and dividing this by the fit asymmetry. This percentage is taken to be the stopping fraction of muons in the pressure cell. Here we see that the simulations match the momentum the experiment was aligned on, at 100MeV/c. The absolute values of the simulations do not match the experimental results completely. This could be due to incorrect initial asymmetries in the sample stopping fraction calculations, or an incorrect spread on the collimation. However, as long as the shape (trend) of the results match the simulations it is clearly reliable enough for alignment and background extraction. 

\begin{figure}[ht]
\includegraphics[width=23pc]{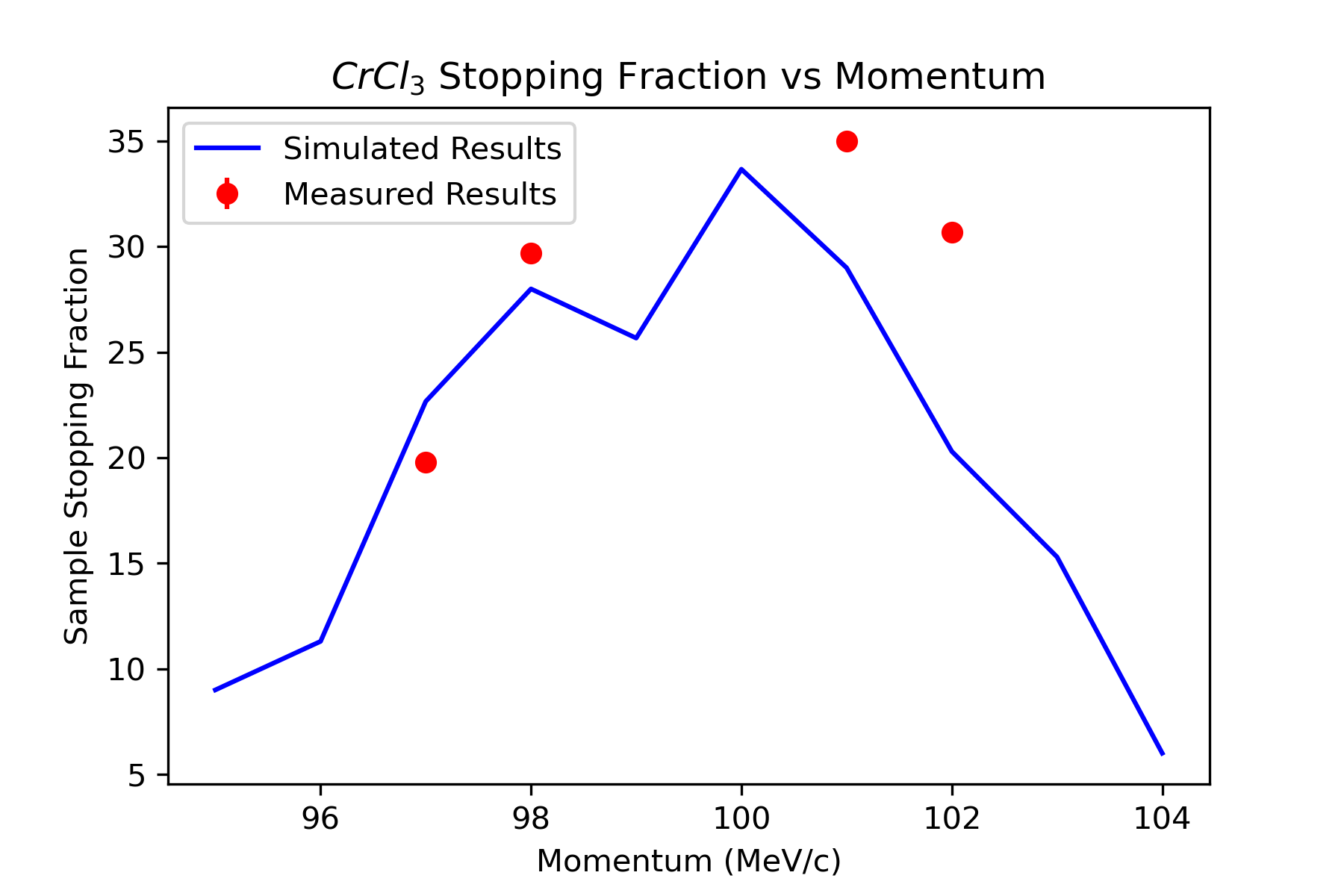}\hspace{0pc}%
\begin{minipage}[b]{14pc}\caption{\label{StoppingPC}The alignment measurements for CrCl$_3$ plotted alongside the simulation results. The sample was aligned using 100MeV/c, which the simulation has shown to be the best momentum to use for the highest stopping percentage.}
\end{minipage}
\end{figure}

\section{Software GUI}\label{StopGui}
\begin{figure}[ht]
  \begin{center}
    \includegraphics[scale=0.55]{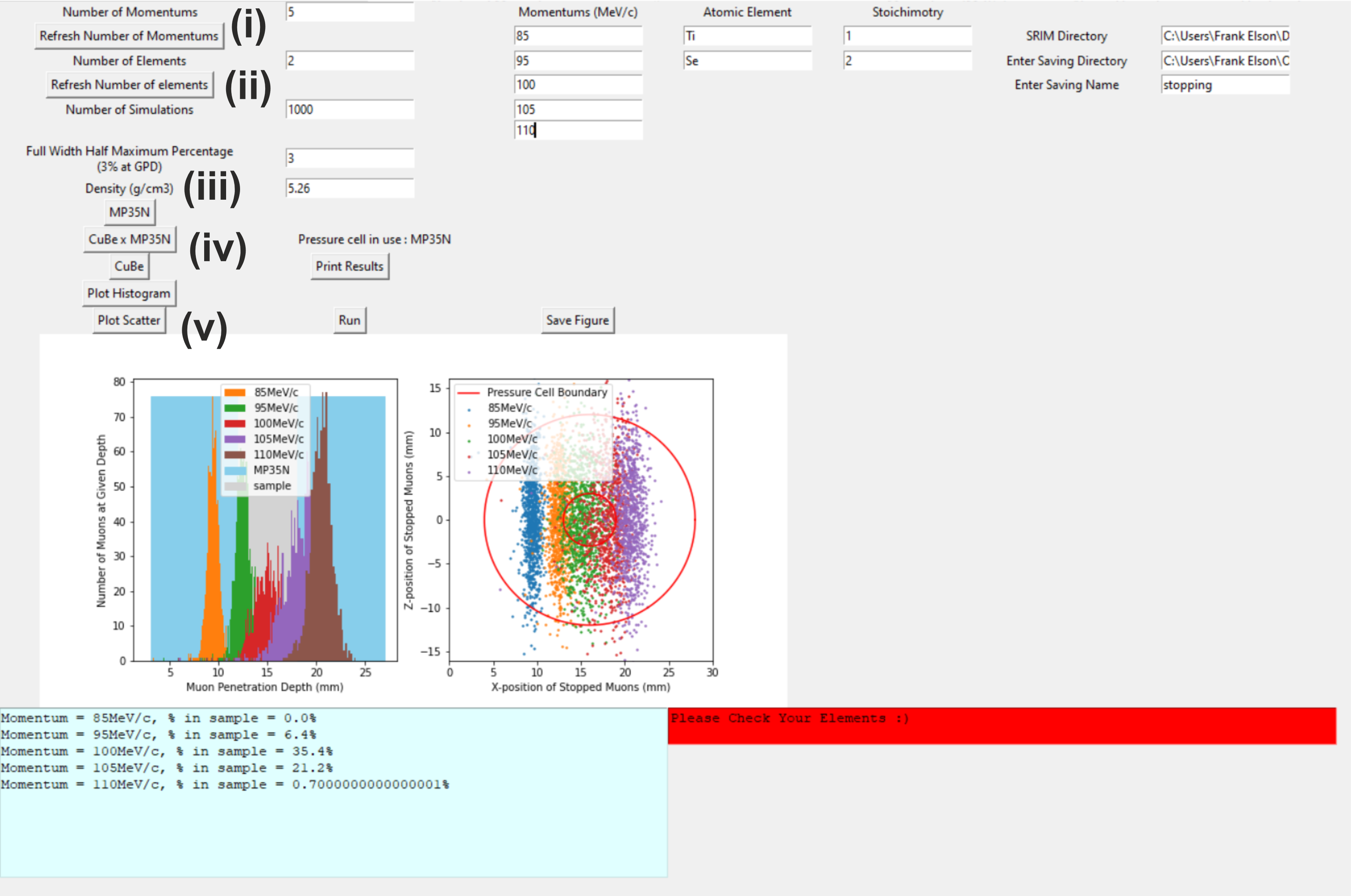}
  \end{center}
  \caption{An example of how the Graphical User Interface (GUI) has been used to calculate the stopping percentages of muons in a sample of TiSe$_2$ at momentums of 85 MeV/c, 95 MeV/c, 100 MeV/c, 105 MeV/c and 110 MeV/c, using the completely MP35N pressure cell. The numbering relates to the list in Section~\Ref{StopGui}.}
  \label{GUI_Whole}
\end{figure}

The interface for running this software (see Fig.~\Ref{GUI_Whole}) has been set up to be user friendly. The goal is to both aid in the alignment process as well as to support data analysis via the calculations of the fractional contributions to the signal. The steps for using the graphical user interface (GUI) are the following (also noted in Fig.~\Ref{GUI_Whole}):

\begin{enumerate}
    \item Enter the desired number of energies you want to simulate in \emph{Number of Momenta}, and click \emph{Refresh Number of Momenta}. This should create a number of text boxes under \emph{Momenta (MeV/c)}, where you fill in the momenta of interest.
    \item Enter the number of elements in your sample in \emph{Number of elements} and click \emph{Refresh Number of Elements}. You should then fill out the element (Note that this is the chemical symbol with the correct capitalised letters) and its associated stoichiomtery. 
    \item Fill out the sample density, the number of simulations, the full width half maximum and also the directory that the SRIM application is on your machine.
    \item Finally, select which pressure cell you would like to use for the simulations and then hit \emph{Run}. Depending on the number of simulations and momenta you have chosen, this could take different amount of time. 
    \item When simulation is done, you can plot the data using \emph{Plot Scatter} or \emph{Plot Histogram}, or you can print your results so that you can obtain the exact positions and also the stopping fractions.
\end{enumerate}

As can be seen from the simulations (graphs in Fig.~\Ref{GUI_Whole}), we can obtain the crucial information about all the muons in the simulations, both collectively and as individual muons. We can also see that for higher momenta the spread of muons parallel to the muon momentum direction is larger than the lower momentum muons. This is because when a higher momentum is selected, these simulations will contain a larger spread of momenta, hence, the stopping range will be larger as well. Additionally, the spread perpendicular to the muon momentum is also increased due to larger scattering effect. Such scattering also affects the simulation time, i.e. as the input momentum is increased, the time is takes to complete the given number of simulations increases also due to the greater scattering. At the point when the energy is great enough that some of the muons will pass through the entire pressure cell, the simulation time reaches a maximum. As the momentum is increased beyond this point, the whole process will take less and less time until all the muons pass through the pressure cell (and then there is nothing more to simulate).

\section{Conclusion and Outlook}
We have developed a user friendly tool that allows for more accurate simulations of the stopping fractions of muons for high-pressure experiments. The options include all the different types of pressure cells provided by the GPD beam line at the Paul Scherrer Institute (PSI) at any given muon momentum using the Variox-Heliox cryostat. The Gaussian distribution of energies in the simulated beam better represents the $\mu$E1 beam line when compared to the single energies of the standard TRIM simulation. The GUI that has been made make it easy for any user to find the stopping percentage of muons in their sample prior to their experiment. This will aid in the alignment process as well as allow the users to extract the background from the collected data. By making this tool available to users it can both help writing their proposals via a feasibility check of proposed experiment depending on how strong their sample response will be. Further, it can also help to help the fitting process via better estimate of signal fractions. We think that this will help solving a key issue faced in high-pressure research in the $\mu^+$SR community, and therefore push the field forward. 

In the future, we would like to further improve this toolbox by extending pySRIM beyond the ability to only make slabs of material. This will clearly lead to an even more accurate representation of the beam setup at GPD. We would also like to implement the possibility to select different sample environments. This includes the different cryostats available at GPD, as well as the new sample holders with different dimensions and thicknesses. 
Additionally, extending this to surface muon beamlines would also be useful. As stated previously, these beamlines have less of an issue with extracting the sample response. However, this can still be a useful tool as you could gain accurate sopping fractions for muons in layers of material, for example. 
Furthermore, other simulation softwares, such as Geant4, would be useful to implement alongside this in a similar fashion. This way we can get different perspectives, and perhaps more accurate results. 

\section{Acknowledgments}
This research is funded by the Swedish Foundation for Strategic Research (SSF) within the Swedish national graduate school in neutron scattering (SwedNess), as well as the Swedish Research Council VR (Dnr. 2021-06157), and the Carl Tryggers Foundation for Scientific Research (CTS-18:272). Y.S is funded by the Swedish Research Council (VR) through a Starting Grant (Dnr. 2017-05078). Y.S. acknowledges funding from the Area of Advance-Material Sciences from Chalmers University of Technology. G.S. acknowledges funding from the Chalmers X-Ray and Neutron Initiatives (CHANS) grant and European Union’s Horizon 2020 research and innovation program under the Marie Skłodowska-Curie grant agreement No 884104 (PSI-FELLOW-III-3i). We thank Matthias Elender and Andreas Suter for valuable discussions. We also thank Ola Kenji Forslun, Yuqing Ge, Gaia De Berardino and Konstantinos Papadopoulos for the CrCl$_3$ alignment data.

\bibliographystyle{unsrt}
\newcommand{\newblock}{}
%\section*{References}

\end{document}